\documentclass[final]{svjour2}
\usepackage{graphicx}
\usepackage{rotating}
\usepackage{amssymb}
\usepackage{mathptmx}
\usepackage[numbers]{natbib}
\makeatletter
\journalname{Journal of Low Temperature Physics}
\bibpunct{}{}{,}{s}{}{,}

\begin{document}
\bibliographystyle{spphys} 
\setcitestyle{numbers,square} 

\newcommand{\hdblarrow}{H\makebox[0.9ex][l]{$\downdownarrows$}-}
\title{Three-nucleon forces and superfluidity in neutron matter}

\author{P. Papakonstantinou$^1$ \and J.W. Clark$^2$}

\institute{% 
$^1$Rare Isotope Science Project, Institute for Basic Science,\\ Daejeon 34047, Republic of Korea\\ 
\email{ppapakon@ibs.re.kr}
%}
%\institute{
\\
$^2$McDonnell Center for the Space Sciences and Department of Physics, Washington University, \\ St. Louis, MO 63130, USA
\\
\email{jwc@wuphys.wustl.edu}}

\maketitle

\begin{abstract}

The existence of superfluidity of the neutron component in the core 
of a neutron star, associated specifically with triplet P-wave pairing, 
is currently an open question that is central to interpretation 
of the observed cooling curves and other neutron-star observables.
{\it Ab initio} theoretical calculations aimed at resolving this
issue face unique challenges in the relevant high-density domain,
which reaches beyond the saturation density of symmetrical nuclear 
matter.  These issues include uncertainties in the three-nucleon (3N)
interaction and in the effects of strong short-range correlations -- 
and more generally of in-medium modification of nucleonic self-energies 
and interactions.  A survey of existing solutions to the gap equations 
in the triplet channel shows that the separate or combined impacts of 
3N forces, coupled channels, and mass renormalization range from 
moderate to strong to devastating, thus motivating a detailed analysis 
of the competing effects. In the present work we track the effects of 
the 3N force and in-medium modifications in the representative case of 
the $^3P_2$ channel, based on the Argonne V18 two-nucleon (2N) interaction 
supplemented by 3N interactions of the Urbana IX family.  Sensitivity of 
the results to the input interaction is clearly demonstrated, while 
consistency issues arise with respect to the simultaneous treatment of 
3N forces and in-medium effects.  We consider this pilot study as the 
first step towards a systematic and comprehensive exploration of 
coupled-channel $^3P\,F_2$ pairing using a broad range of 2N and 3N 
interactions from the current generation of refined semi-phenomenological 
models and models derived from chiral effective field theory.


\end{abstract}

\section{Introduction}

Nucleonic systems demonstrate superfluid and superconducting properties 
in a variety of density and composition regimes~\cite{DeH2003}. Odd-even 
staggering served historically as the first clue to the presence of 
superfluidity in finite nuclei, while rotational spectra exhibit the 
effect of superfluidity on the nuclear moments of inertia.  The formation 
and properties of neutron halos in light drip-line nuclei, currently under 
investigation theoretically and in modern radioactive-beam facilities, are 
also influenced by pairing.  Neutron stars serve as rich laboratories for 
nuclear-matter studies across a vast range of densities.  Cooling via 
neutrino emission, glitches, and rotational dynamics are all sensitive to 
the pairing strength in different two-nucleon channels and for different 
isospin compositions~\cite{Lei2015,EEH1996}.  In particular, the presence 
of nucleonic superfluidity and the magnitude of the pairing gap in the 
homogeneous hadronic phase present in the outer core of a neutron star 
%#
is an open question, whose answer is pivotal for interpretation of 
%#
the observed cooling data \cite{Sch1997,Page2014,Gez2014}.  

Unlike electronic systems in terrestrial solids, where the pairing mechanism 
depends on the presence of an ionic lattice, the attractive force generating 
pairing in nuclear systems comes directly from the interactions between 
and among the nucleons themselves.  These interactions, possessing both 
attractive and repulsive components, exhibit rich operatorial structure
expressing complicated dependence on space, spin, and isospin, giving rise 
to diverse and subtle pairing phenomena.  At the most basic level, we may 
distinguish between isoscalar pairing, active in isospin-symmetric nuclei 
in the deuteron channel, and isovector pairing, especially relevant in 
halo nuclei and neutron-star matter.  Next we may consider the partial-wave 
decomposition of the nucleon-nucleon interaction, with different angular 
momenta coming into play becoming more important at different relative 
energies, corresponding to different nucleonic densities.  In particular, 
one may infer that singlet $^1S_0$ pairing of neutrons is active at 
subsaturation densities, i.e., in nuclear haloes and in the neutron fluid 
interpenetrating the rigid pasta-like structures that may be formed by the 
neutron-rich nuclei in the inner crust of a neutron star, while neutron 
pairing in higher partial waves, if present, affects only the dense hadronic 
liquid at greater depth in in the star's outer core.  In the latter case, 
the tensor interaction couples different angular-momentum channels, and 
anisotropic pairing gaps cannot be excluded, 
%# 
specifically in the triplet $^3P\,F_2$ state. 
%#
Proton pairing in the $^1S_0$ channel is expected to occur in this same 
density region, where the partial density of the sparse proton admixture 
required for $\beta$-stability will be similar to that of neutrons in 
the inner crust.

There exist many calculations of the pairing gap in the singlet $S$-wave 
channel (for recent reviews and comparisons, see 
\cite{Gez2014,50yrs,PMM2016X}, and references cited therein). 
While different calculational methods have yielded a considerable 
spread of predicted values, there is a growing consensus that in pure 
neutron matter the gap value should reach a maximum of about 2 MeV at a 
Fermi momentum somewhat lower than 1~fm$^{-1}$.  Results from many different 
approaches are compiled, for example, in Fig.~6 of Ref.~\cite{PMM2016X}.  
In view of the extreme (nominally exponential) sensitivity of the pairing 
gap to the pairing interaction and the density of single-fermion states, 
the lack of quantitative agreement of the different microscopic approaches 
is not surprising. 

In this contribution we focus on issues of the existence and intensity
of pairing as it might occur in the liquid hadronic outer core of a neutron 
star, at densities ranging up to several times the saturation density of 
symmetrical nuclear matter, where triplet neutron pairing is most likely 
to occur.  The theoretical challenges encountered under the extreme conditions 
in play abound and uncertainties accumulate.  At such high densities, the 
3N force, which reflects the composite nature of nucleons and is poorly 
constrained, cannot be neglected.  Also inadequately controlled are 
significant short-range correlations and self-energy effects, whose 
treatment involves approximations of variable reliability.   

%#
It should be noted that the effect of the 3N force on the singlet-$S$ 
pairing gap has been found to be quite minor, owing to the much lower
density at which this gap reaches its peak value, occurring at a
small fraction ($\sim 1/8-1/10$) of the saturation density of symmetrical 
nuclear matter.  In theoretical calculations, this effect would be masked 
by other uncertainties involving medium-polarization and self-energy
corrections~\cite{ZSZ2004,Yuan2007}.   
%#

%#
To set the context for our study, let us consider briefly the results of 
three previous caculations of the isospin $T=1$ triplet gap at neutron-star 
densities that include a 3N interaction.  For relevant earlier work on 
the $^3P_2$ pairing problem and its coupled-channel extension $^3P\,F_2$, 
both theoretical and computational, see Refs.~\cite{Tam1970,Tak1971,Tak727384,Amu1985,Bal1992,Tak1993,Bal1998,Kho1998,Kho2001a,Kho2001b,Zve2003}.   

Contributing to the long-term effort of the Catania group on nucleonic
pairing, W.~Zuo et al.~\cite{ZCL2008} employed the Argonne $v_{18}$ (AV18) 
two-nucleon (2N) potential along with a modified Urbana IX (UIX) 3N 
interaction \cite{Pud1997} in a study of $^3P\,3F_2$ pairing in pure 
neutron matter.  The parameters of the original UIX 3N potential were 
readjusted in order to obtain realistic results for saturation of 
symmetrical nuclear matter, as calculated within Brueckner-Hartree-Fock 
(BHF) theory~\cite{Zho2004}.  The modified 3N potential is far less 
repulsive at short distances than the original version.  The latter was 
developed by the Argonne-Urbana group with parameters which, in conjunction 
with the AV18 interaction, (i) reproduce the empirical saturation properties 
of symmetrical nuclear matter as calculated in the single-operator-chain 
version of variational many-body theory, while also (ii) yielding 
satisfactory consistency with three-nucleon binding.  Inclusion, by Zuo 
et al., of an additional term in the pairing interaction corresponding to 
the modified UIX 3N interaction resulted in a subtantial enhancement of the 
gap's maximal value over that obtained for the AV18 2N potential alone, 
namely about 0.7~MeV, to almost 1.2~MeV, at $k_F\approx 2.2$fm$^{-1}$.  This 
result was obtained using a free single-particle spectrum in the denominator 
of the gap equation.  Upon replacing the free spectrum by its 
Brueckner-Hartree-Fock counterpart so as to include some in-medium 
effects (``correlations''), the gap was reduced to about 0.5 MeV.

In a more recent study along similar lines, J.~M.~Dong et al.~\cite{DLZ2013} 
utilized the Bonn B 2N potential, together with a meson-exchange 3N 
potential with the same meson parameters as Bonn B~\cite{LLS2008}.  
They found a strong enhancement of the gap and a shift of its peak 
position versus $k_F$ when the 3N force was included.  The result with 
the 3N force in effect showed a maximal value of the gap of about
0.6~MeV at $k_F\approx 2.3$fm$^{-1}$.  When the effect of correlations 
was estimated in terms of the so-called $Z-$factor, which quantifies 
the depletion of the fermion occupation number around the Fermi surface, 
the gap was reduced by an order of magnitude to an insignificant maximal 
value of about 50~keV at $k_F\approx 1.8$fm$^{-1}$. 

S.~Maurizio et al.~\cite{MHF2014} considered chiral 2N and 3N potentials, 
including a high-precision 2N potential developed at next-to-next-to leading 
order in the chiral expansion.  In a representative example, taking 
account of the 3N force resulted in a modest enhancement of the gap by 
some 20\%, yielding a maximal gap of 1.15~MeV, close to the uncorrelated
result of Ref.~\cite{ZCL2008}.  Inclusion of in-medium modifications of
the effective mass reduced the gap by approximately one third to a 
maximal value about 0.35~MeV at $k_F\approx 2.2$fm$^{-1}$. 

Based on the above survey, we infer that the combined effect of 
3N forces and correlations on the results of calculations 
of the triplet pairing gap in neutron matter ranges from moderate to 
strong to devastating, depending on the precise inputs.  Morover, 
the effects of the three-nucleon force on one hand and correlations on
the other can be both large and opposite.

%#
Pointing to selected earlier investigations, it becomes apparent that the 
input 3N interaction itself, even if its choice must somehow be plausible, 
can quite readily either make or break a triplet gap found to be present
for the 2N interaction alone.  A microscopic 3N 
%#
potential obeying certain consistency requirements with respect to the AV18 
potential has been shown~\cite{GLM1989} to produce a weaker triplet-pairing 
gap in stellar matter than obtained by Zuo et al.\ \cite{ZCL2008} using the 
modified UIX~interaction. On the other hand, use of the original UIX 
%#
parameterization emphatically kills the gap at least in the uncoupled 
$^3P_2$ channel~\cite{Yuan2007}.  (We will follow up on this latter finding, 
in considerable detail.)  It is important to note that all three of these 
versions of the 3N potential, when combined with the AV18 2N interaction, 
reproduce the empirical saturation properties of nuclear matter within 
specific (and competing) many-body methods.  Accordingly, such disagreement 
in the solutions of triplet gap equations raises issues of consistency 
for the many-body theories {\it vis {\`a} vis} the interactions involved.

In the present work, the sensitivity of the results to the input interaction, 
sometimes dramatic, will be clearly demonstrated.  Thus, one's control 
of the approximations involved, or rather the lack thereof, will be 
revealed.  We introduce two novelties relative to the prevalent treatments 
of gap equations in the presence of a 3N interaction, which were proposed and 
originally implemented in Ref.~\cite{Yuan2007}.  First, we avoid the usual 
approximation of averaging the 3N interaction over the third (unpaired) 
particle to produce a 
density-dependent
effective 2N interaction.  Rather, we introduce the explicit 3N interaction 
when constructing the two-body pairing-interaction matrix elements in 
momentum space, for insertion into the gap equation.  (This will introduce 
a term in the pairing matrix elements having an additional momentum 
integration over the unpaired neutron.) Second, along with results for 
the modified UIX potential, we obtain results with the original 
parameterization of the UIX potential which yields a proper equation 
of state within variational chain-summation methods~\cite{APR1998} and 
predicts a neutron star mass-radius relation which is consistent 
within observational constraints~\cite{Dem2010,Ant2013,SLB2010}. 

The present exploration focuses on pure neutron matter and on the $^3P_2$ 
channel.  The coupled-channel problem will be addressed in future work.
The purpose at present is not to provide the most realistic predictions, 
but instead to trace the sensitivity of the results to the most basic 
inputs.  As will be seen, already within this simplified model system 
the inclusion and parameterization of the 3N force is of supreme 
importance, and very basic questions arise for quantum many-body theory:
What does a ``consistent'' treatment of the 3N interaction entail? 
Given that the 3N interaction is already in a sense an in-medium effect, 
what additional in-medium effects must be considered and how?  Such 
questions are especially pertinent in the era of {\it ab initio} nuclear 
structure, as pairing becomes a testing ground for modern chiral 
interactions~\cite{MHF2014}.

%#
%#      
The paper is organized as follows. In Sec.~\ref{Sec:Eqs} we sketch
the theoretical basis for our study, namely the BCS approach extended
to triplet pairing, applied in angle-averaged approximation, and show 
baseline results for the triplet $^3P_2$ pairing gap. Importantly,
we explain our treatment of the 3N interaction and the formulae 
pertinent to the implementation of the UIX potential.  Our main
results are presented in Sec.~\ref{Sec:NNN}.  In particular, 
Sec.~\ref{Sec:NNNsc} tracks the influence of the attractive 
part and the phenomenological repulsive part of the Urbana potential 
by varying the coupling constants.  At first, kinetic energies
are used for single-particle energies with the bare neutron
mass, and no correlations or in-medium effects are considered, 
beyond those implicit at the BCS level.  In Sec.~\ref{Sec:NNNmed} 
we discuss results based on an effective nucleon mass.  Also in 
Sec.~\ref{Sec:NNNmed} we introduce a defect function to parameterize 
the correlations experienced by the third unpaired nucleon.  
In Sec.~\ref{Sec:NNNV} the results are analyzed in terms of 
the overall nucleon-nucleon pairing potential near the Fermi 
surface.  Conclusions are drawn and perspectives are given 
in Sec.~\ref{Sec:End}.

\section{BCS equations for three interacting nucleons and baseline results\label{Sec:Eqs}}

\subsection{Basic gap equations and numerical solution\label{Sec:Eqs:Eqs}} 

%#
Derivations of the BCS gap equations for isospin $T=1$ triplet pairing can 
be found in several sources; for example, see Refs.~\cite{Amu1985,Tak1993}).
A suitable generalization of the BCS superfluid trial ground state for 
singlet pairing to the more complicated case of triplet pairing can
take the form
\begin{equation}
| \Phi_s \rangle = {\cal N}^{-1}\prod_{\sigma\sigma'{\vec k},k_x>0} 
\left(1 + g_{\sigma \sigma'}({\vec k})
a_{{\vec k}\sigma}^\dagger a_{-{\vec k}\sigma}^\dagger \right) | 0 \rangle,
\end{equation}
where $|0\rangle$ is the quasiparticle vacuum and $\cal N$ is a normalization 
constant. The factor $g_{\sigma,\sigma'}({\vec k})$, though a matrix, plays 
the role of a variational function, analogous to $h({\vec k})$ of the original 
BCS derivation \cite{BCS}.  This state does not conserve the particle number 
$N=\langle\hat{N}\rangle=\langle \sum_{\vec{k}\sigma} a_{\vec{k}\sigma}^{\dagger} 
a_{\vec{k}\sigma}\rangle$. Consequently $g_{\sigma \sigma'}({\vec k})$ is 
to be determined by functional minimalization of the expectation 
$\langle \hat{H}-\mu\hat{N}\rangle$ subject to the constraint of a constant 
quasiparticle density. Here $\hat{H}$ is the Hamiltonian and $\mu$ is 
a Lagrange multiplier interpreted as the chemical potential.  The resulting 
matrix analog of the BCS gap equation can be reduced by partial-wave 
analysis, leading to a set of gap equations in general coupled with one 
another, for gap components $\Delta^{LSJM}_{\vec{k}}$ belonging to specific 
spin/angular-momentum channels.  Such partial-wave gap equations have 
been derived and studied extensively by Takatsuka and Tamagaki \cite{Tak1993}, 
and the equations for pairing in the $^3P\,F_2$ coupled-channel case have 
been analyzed in detail more recently by Khodel et al.~\cite{Kho2001b,Zve2003} 
based on a separation algorithm for their solution \cite {Kho1996,Kho2001a}.

At present we are concerned primarily with the uncoupled triplet channel 
$^3P_2$.  Upon averaging over angles and five components corresponding
to the allowed $M$ projections for $J=2$, the angle-averaged $^3P_2$ gap 
function, henceforth denoted simply as $\Delta(k)$, is found 
(cf.~\cite{Yuan2007}) to obey an equation of the same form as that for 
the $^1S_0$ gap, 
\begin{equation} 
\Delta (k)  = -\frac{1}{\pi} \int_0^{\infty} 
 dk' {k'}^2 \frac{V(k,k')\Delta(k')}{\sqrt{[ (\epsilon_{\vec{k'}} 
- \mu ) ]^2 + [\Delta(k')]^2}}, 
\label{Eq:GapEq} 
\end{equation} 
where $V(k,k')$ is the central part of the potential belonging to the given 
multipolarity, as represented in momentum space. 
Invoking the separation method~\cite{Kho1996,Kho2001a,Yuan2007}, 
numerical solution for a given value of Fermi momentum $k_F$ is facilitated 
by introducing certain functions $f(k)$, $\chi(k)$, and $W(k,k')$ through 
the definitions
\begin{equation} 
\Delta (k) \equiv \Delta_F f(k), \qquad   \Delta_F \equiv \Delta (k_F),  
\end{equation} 
\begin{equation} 
V_{F}\equiv V(k_F,k_F), \qquad  \chi (k) \equiv V(k,k_F)/V_{F}, 
\end{equation} 
\begin{equation} 
W(k,k') \equiv V(k,k') - V_{F}\chi(k)\chi(k'). 
\end{equation} 
Importantly, we note the property $W(k,k_F)=W(k_F,k)=0$, which allows
the gap equation for given $k_F$ to be recast as the following system
of two coupled equations:
\begin{equation}
f(k) + \frac{1}{\pi} \int_0^{\infty} dk' {k'}^2 
\frac{W(k,k')f(k')}{\sqrt{[ (\epsilon_{\vec{k'}}-\mu ) ]^2+[\Delta_Ff(k')]^2}}, 
\label{Eq:Delta1} 
\end{equation}  
\begin{equation} 
\frac{V_F}{\pi} \int_0^{\infty} dk' k'^2 \frac{\chi(k')f(k')}
{\sqrt{[ (\epsilon_{\vec{k'}} - \mu ) ]^2 + [\Delta_Ff(k')]^2}}    =  -1 . 
\label{Eq:Delta2}
\end{equation} 
The unknowns are $\Delta_F$ and $f(k)$.  In fact, the single-particle energies 
$\epsilon_{\vec k}$ and chemical potential $\mu$ are also unknown and should 
be determined self-consistently, greatly complicating the problem.  Instead
we shall assume that $\epsilon (k)$ and $\mu$ are known within some 
approximation, as considered in Sec.~\ref{Sec:Eqs:InMedium}.  Solution
then becomes rather straightforward: In an initial step, we set 
$\Delta_F f(k)=\delta_0$, a small constant, in the denominator of 
Eq.~(\ref{Eq:Delta1}).  This equation then becomes a standard Fredholm 
equation of the second kind with respect to $f(k)$ and can be solved 
using standard routines.  Given $f(k)$, the root $\Delta_F$ of 
Eq.~(\ref{Eq:Delta2}) can be found, for example, through bisection. 
The new $\Delta_F$ can be fed back into Eq.~(\ref{Eq:Delta1}). 
Consistency is achieved quite fast in practice.  Also, as a
practical matter, the potential $V(k,k')$ need not be calculated on 
the fly. Rather, we store its values on a $k$-space mesh, in a file 
from which they can be read each time calculations need to be executed. 
  
\subsection{Three-nucleon gap equations} 

The derivation of the 3N gap equation is detailed in Ref.~\cite{Yuan2007}. 
Here we outline the basic steps and assumptions.  An important assumption 
is again the angle-average approximation, a partial-wave decomposition 
being implied.  Here we consider only the triplet $^3P_2$ channel, 
without coupling to the $^3F_2$ channel.  

%#BEGINNING OF REPLACEMENT - new paragraph
The angle-averaging treatment has been commonly applied in previous numerical 
studies of triplet-$P$ pairing in nucleonic media, as in 
Refs.~\cite{Bal1998,Kho2001a,ZCL2008}.  It is well justified for two 
reasons.  First, there exists a striking near-degeneracy of the energies 
of the five magnetic substates involved in the uncoupled $^3P_2$ problem, 
irrespective of the details of the interaction, i.e., such behavior 
is universal.  This property, having a sound theoretical basis, was 
discovered by Takatsuka and Tamagaki \cite{Tak1971,Tak1993} and 
validated unambiguously in more recent analyses \cite{Kho1998,Kho2001a},
that exploit the separation method for solving coupled gap equations.  
(The near-degeneracy is lifted somewhat by the tensor force acting in 
the $3P\,F_2$ coupled-channel case, where there exist thirteen real 
solutions, another universal property \cite{Kho2001b,Zve2003}).  The 
second reason is strategic: the angle-averaged gap is the quantity 
of most immediate importance in  the astrophysical implications of 
the existence and strength of triplet-$P$ pairing for neutron-star cooling.

Accordingly, we will be working with the 2N pairing potential $V(k,k')$ and 
an additional 3N pairing potential $V_3(k,k',k'')$, without angular 
dependencies.%END OF REPLACEMENT
Thus two nucleons form a Cooper pair in the $^3P_2$ 
channel and a spectator nucleon can interact with them.  Only $^1S_0$ 
interactions are assumed between the spectator and the other particles, 
higher partial waves being ignored at present, in an approximation we 
adopted from Ref.~\cite{ZSZ2004} that may require improvement at high 
densities.  It turns out that the gap obeys Eq.~(\ref{Eq:GapEq}) just as 
in the 2N case, with the difference that the two-body potential $V(k,k')$ 
entering the equation receives an additive contribution from the 3N 
potential that reads~\cite{Yuan2007} 
\begin{eqnarray} 
V^{(3)} (k,k') &=&  \frac{1}{\pi} \int_0^{\infty} dk'' k''^2 V_3(k,k',k'') 
\nonumber \\ 
  & & \times 
\int_{-1}^{1} dx \left[ 
1 - \frac{\epsilon_{k''} - \mu }{\sqrt{(\epsilon_{k''}-\mu )^2 + \frac{1+3x^2}{16\pi }[\Delta(k'')]^2}} 
\right] 
\\ 
& \approx & \frac{4}{\pi} \int_0^{k_F} dk'' k''^2 V_3(k,k',k'')  
\equiv V^{(3)} (k,k';k_F). 
\label{Eq:V3k} 
\end{eqnarray} 
(The indicated approximation, which is quite accurate, takes advantage of the
smallness of the triplet gap relative to the Fermi energy corresponding
to relevant $k_F$ values \cite{Yuan2007}.)

Finally, we give the expression for $V_3(k,k',k'')$ in the special case of 
an Urbana-type potential, consisting of an attractive two-pion exchange 
term of strength $A_{2\pi}$ and a phenomenological repulsion term of 
strength $U_0$~\cite{Yuan2007}:  
\begin{eqnarray} 
V_3(k,k',k'') &=& 
\frac{1}{2kk'} \int d\tilde{r}_1 d\tilde{r}_2dx \,\, k\tilde{r}_1j_1(k\tilde{r}_1) \left\{  
k'\tilde{r}_1j_1(k'\tilde{r}_1) \tilde{r}_2^2 
\right. 
\nonumber 
\\ 
 & & \quad \times \{ 2A_{2\pi} [-6(Y_1Y_2+Y_3Y_1) + \frac{12}{5}(T_1Y_2+Y_3T_1) + 18 Y_2Y_3] 
 \nonumber \\ 
 & & \quad \quad \quad + U_0 [T_1^2T_2^2 + T_2^2T_3^2 + T_3^2 T_1^2 ] \} 
\nonumber \\ 
 & & 
+  (1+3x^2) \tilde{r}_1\tilde{r}_2j_0(k''\tilde{r}_3)  k'\tilde{r}_2j_1(k'\tilde{r}_2) 
\nonumber \\ 
& & \quad \times 
\{   2A_{2\pi} [2Y_1Y_2 -6(Y_2Y_3+Y_3Y_1) +\frac{12}{5}(T_2Y_3+Y_3T_1) 
 \nonumber \\ 
 & & \qquad\qquad\qquad  -\frac{4}{5}(Y_1T_2+T_1Y_2) + \frac{8}{25}T_1T_2 ] 
\nonumber \\ 
& & \quad \quad \quad \left. 
+ U_0 [T_1^2T_2^2 + T_2^2T_3^2 + T_3^2 T_1^2 ] \}  
\right\} ,
\label{Eq:urb44} 
\end{eqnarray} 
where $j_{\ell}(x)$ is a spherical Bessel function and we have defined  
\begin{equation} 
\tilde{r}_1 = |\vec{r}_1-\vec{r}_2| 
\, \, , \, \,  
\tilde{r}_2 = |\vec{r}_2-\vec{r}_3| 
\, \, , \, \,   
\tilde{r}_3 = |\vec{r}_3-\vec{r}_1| = \sqrt{\tilde{r}_1^2 + \tilde{r}_2^2 - 2\tilde{r}_1\tilde{r}_2x}
\label{Eq:tilder} 
\end{equation} 
in terms of the three particle coordinates. The radial and tensor Yukawa functions are given by 
\begin{equation} 
Y_i = \frac{\mathrm{e}^{-\xi \tilde{r}_i }}{\xi \tilde{r}_i} 
\left( 1 - \mathrm{e}^{-C \tilde{r}_i^2 }\right) 
\quad , \quad 
T_i = \left(  1 + \frac{3}{\xi \tilde{r}_i}  
+ \frac{3}{\xi^2\tilde{r}_i^2} \right) \frac{\mathrm{e}^{-\xi \tilde{r}_i }}
{\xi \tilde{r}_i} \left( 1 - \mathrm{e}^{-C \tilde{r}_i^2 }\right)^2,  
\end{equation} 
with parameters $\xi=m_{\pi}c/\hbar$ (involving the average pion mass 
$m_{\pi}$) and $C=2.1$~fm$^{-2}$. 

In practice we split $V^{(3)}(k,k';k_F)$ into the two contributions 
coming from its attractive and repulsive parts, 
\begin{equation} 
V^{(3)}(k,k';k_F) = V^{(3)}_{A_{2\pi}}(k,k';k_F)  +  V^{(3)}_{U_0}(k,k';k_F) 
\label{Eq:V3split} 
\end{equation} 
in an obvious notation.  For each $k_F$ value of interest, 
$V^{(3)}_{A_{2\pi}}(k,k')$ and $V^{(3)}_{U_0}(k,k')$ are calculated by 
numerical integration, and rows of $k,\,k',\,V^{(3)}_{A_{2\pi}}(k,k'),
\,V^{(3)}_{U_0}(k,k')$ values are stored in a $k_F$-specific file.  
Eventually the two additive contributions can be read and utilized 
as separate entities.  Thus they can easily be scaled at will during calculations. 

The calculation of $V^{(3)}(k,k';k_F)$ is very time-consuming and 
so far we have performed it for just a few representative values of 
$k_F$.  For those values we were then able to calculate $\Delta_F$. 
Fortunately, the functional behaviour of $\Delta_F(k_F)$ can be 
accurately fitted to a Gaussian function over most of its range; 
the feature is exploited in plotting our results as a function of 
$k_F$ in Fig.~\ref{Fig:3N}. 

\subsection{In-medium effects\label{Sec:Eqs:InMedium}} 

As mentioned previously, the single-particle spectrum $\epsilon (\vec{k})$ 
and the chemical potential $\mu$ are assumed known. In principle these
quantities should be derived self-consistently for the given Hamiltonian,
density, and temperature (zero in the present case).  Lacking that, for 
exploratory purposes one may consider a range of approximations based 
on collective experience with similar problems in nuclear many-body 
physics.  The most rudimentary assumption corresponds to the non-interacting 
Fermi gas, 
\begin{equation} 
\epsilon_{\vec{k}} = \frac{\hbar^2}{2m} + {\cal U}, 
\qquad \mu = \epsilon_{\vec{k}_F} 
\equiv \epsilon_F, 
\end{equation} 
where $m$ is the bare nucleon mass and ${\cal U}$ a constant potential.  More 
correctly, we should consider a self-consistent single-particle potential 
${\cal U}_{\vec{k}}$, or to lowest order the BCS equations for quasiparticles 
with an effective nucleon mass $m^{\ast}\neq m$.  In Ref.~\cite{ZCL2008}, 
for example, the neutron effective mass in neutron matter at different 
densities was calculated within a Brueckner-Hartree-Fock approach, 
first assuming a 2N potential only (Argonne $v_{18}$) and then assuming 
an additional 3N interaction.  In the present study we consider several
choices for $m^{\ast}$, including those from Ref.~\cite{ZCL2008}.  

Given the strong short-range repulsions present in the Hamiltonian, 
we should also be concerned with the presence of short-range correlations, 
in particular those between the third, unpaired particle and the paired 
ones in the proper evaluation of $V(k,k',k'')$. (Such correlations between 
the paired particles themselves, induced by the inner repulsion of AV18, are
taken care of to a certain degree in the trial ground state 
itself, sufficient to avoid singular behavior \cite{CMS1959}.)
To account for this specific correlation induced by the 3N interaction, 
we simply introduce a defect function, essentially a Jastrow correlation 
factor of gaussian type, 
\begin{equation} 
g(\vec{r}) = \exp\{ -r^2 / b^2 \} , 
\end{equation} 
which suppresses the contribution of the repulsive component of the 3N 
potential whenever the third, unpaired nucleon comes close to its 
paired neighbors.  The 3N potential $V_3(\vec{r}_1,\vec{r}_2,\vec{r}_3)$ 
in coordinate space is then replaced by 
\begin{equation} 
\tilde{V}_3(\vec{r}_1,\vec{r}_2,\vec{r}_3) = 
V_3(\vec{r}_1,\vec{r}_2,\vec{r}_3) [1-g(\tilde{r}_2)]^2 [1-g(\tilde{r}_3)]^2,
\label{Eq:V3def} 
\end{equation} 
with $\tilde{r}_i$ defined in Eq.~(\ref{Eq:tilder}).  This substitution 
is straightforward to implement in the calculation of the contribution of 
the 3N potential to the gap equation, just by inserting the factor
$[1-g(\tilde{r}_{3})]^2 [1-g(\tilde{r}_{2})]^2$ into the integral in 
Eq.~(\ref{Eq:urb44}).  

In our numerical studies, the values $0.6$ and $0.8$~fm are considered 
for the range $b$ of the correlations, $(k_F,b)$-specific files being 
created for storing $V^{(3)}_{A_{2\pi}}$ and $V^{(3)}_{U_0}$. 

\subsection{Baseline results\label{Sec:Eqs:Baseline}} 

\begin{figure}
\begin{center}
\includegraphics[%
  width=0.55\linewidth,angle=-90,
  keepaspectratio]{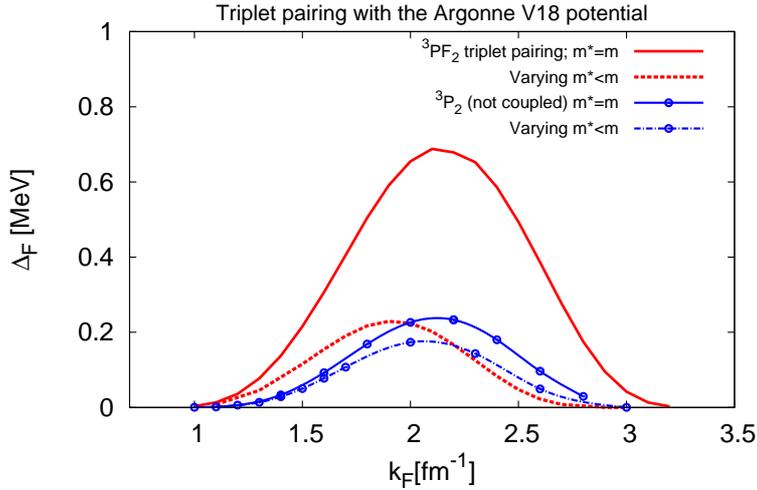}
\end{center}
\caption{Pairing gap in the triplet channel as a function of Fermi momentum 
$k_F$ assuming a bare nucleon mass ({\it solid lines}) or an effective mass 
adopted from Ref.~\cite{ZCL2008} ({\it broken lines}).  Results for the 
coupled $^{3}PF_2$  channel ({\it solid and dashed red}) are compared with 
results for the uncoupled $^3P_2$ channel ({\it solid with points and 
dash-dotted blue}).  (Color figure online.)}
\label{Fig:Coupled}
\end{figure}

In Fig.~\ref{Fig:Coupled} we show results for the Fermi-surface pairing gap 
in the triplet $T=1$ channel as calculated with the AV18 interaction alone. 
Comparison of the two solid lines demonstrates the effect of introducing 
the coupling between the $P$ and $F$ channels on the angle-averaged gap: 
At the peak position ($k_F\approx 2.1$~fm$^{-1}$) the gap is more than 
tripled. In this case we have considered an effective nucleon mass equal 
to the bare mass, $m^{\ast}=m$.  Our results agree with independent 
calculations in the literature, thus validating the numerical implementation. 

If we take into account the in-medium modification of the nucleon mass, 
the effect of the coupling is moderated substantially,
as comparison of the dashed and dot-dashed lines in Fig.~\ref{Fig:Coupled} shows. 
For these calculations the values of the effective mass in neutron matter as a 
function of the Fermi momentum were taken from Ref.~\cite{ZCL2008} 
(two-body case).  

Next we turn to our main subject, the effect of the 3N force combined with 
in-medium effects on the pairing gap. We will consider only the uncoupled 
$^3P_2$ problem at present.  This case will suffice for demonstrating the 
sensitivity of the results to the parameters chosen.  The additional effect 
of $PF$ coupling will be the subject of a future investigation.  

\section{Tracking the effect of the three-nucleon force \label{Sec:NNN}}

\subsection{Scaling the three-nucleon force \label{Sec:NNNsc}} 

First we replicate the results of Ref.~\cite{Yuan2007} for the UIX potential. 
At this point we should again stress that different parameterizations have been 
used in the literature in solving the BCS equations: The original one was 
tuned to the triton binding energy and the saturation point of nuclear 
matter within variational CBF theory.  A modified parameterization was 
employed by the Catania/Lanzhou group, readjusted to describe the saturation 
point within BHF theory~\cite{Zho2004}.  The two sets of parameters are 
given in Table\ref{table}.
\begin{table} 
\begin{center} 
\begin{tabular}{|c|cc|}
\hline 
    & original & modified \\ 
\hline 
$A_{2\pi}$  &  -0.0293  &  -0.0333 \\   
$U_0$   &  0.0048  & 0.00038 \\ 
\hline  
\end{tabular} 
\end{center} 
\caption{Coupling constants, in units of MeV, of the original and modified 
parameterizations of the Urbana IX potential.\label{table} }
\end{table}  
The attractive part of the modified UIX is $14\%$ stronger than the original 
value, while the modified repulsive part is one order of magnitude weaker 
than the original one. Overall, the modified potential is much less repulsive. 

\begin{figure}
\begin{center}
\includegraphics[%
  width=0.55\linewidth,angle=-90, 
  keepaspectratio]{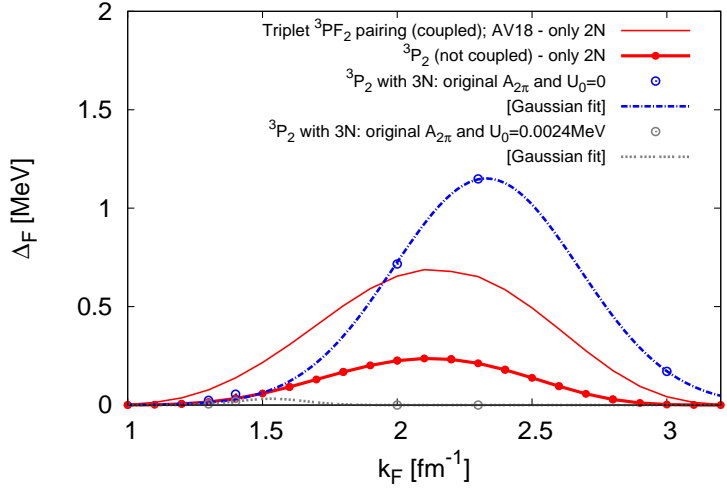}
\end{center}
\caption{Pairing gap in the triplet channel as a function of Fermi momentum 
$k_F$ assuming a bare nucleon mass and different possibilities for the 3N 
force: (i) not included ({\it red solid line with points}), (ii) including the 
original value for the UIX potential $A_{2\pi}=-0.0293$~MeV and setting 
$U_0$ to zero ({\it blue dash-dotted line}) or to $U_0=0.0024$MeV, i.e., 
half its original value ({\it grey dotted line}). The open circles correspond 
to numerical results; dotted and dashed-dotted lines were generated from 
fitting to those results. (Color figure online.)}
\label{Fig:3N}
\end{figure}

In agreement with Ref.~\cite{Yuan2007}, we found that if the original UIX 
force is used, there is no positive solution for the gap: $\Delta_F(k)\equiv 0$.  
The modified UIX interaction does not produce such dramatic behavior.  
Representative results will be discussed below.  It turns out that the 
strength of the repulsive term represented by $U_0$ is a crucial factor.  
In Fig.\ref{Fig:3N}, we show results obtained by neglecting the $U_0$ term 
and compare them with the results with no 3N force present: the gap increases 
strongly.  On the other hand, setting $U_0$ to half its original value, 
namely to $0.0024$ MeV, leads to a very strong reduction of the gap. It 
is thus quite revealing to track the behavior of the gap value as the 
%#JWC
strength of the overall UIX force or separately its repulsive vs. attractive component is varied. 
%#

\begin{figure}
\begin{center}
\includegraphics[%
  width=0.65\linewidth,angle=-90, 
  keepaspectratio]{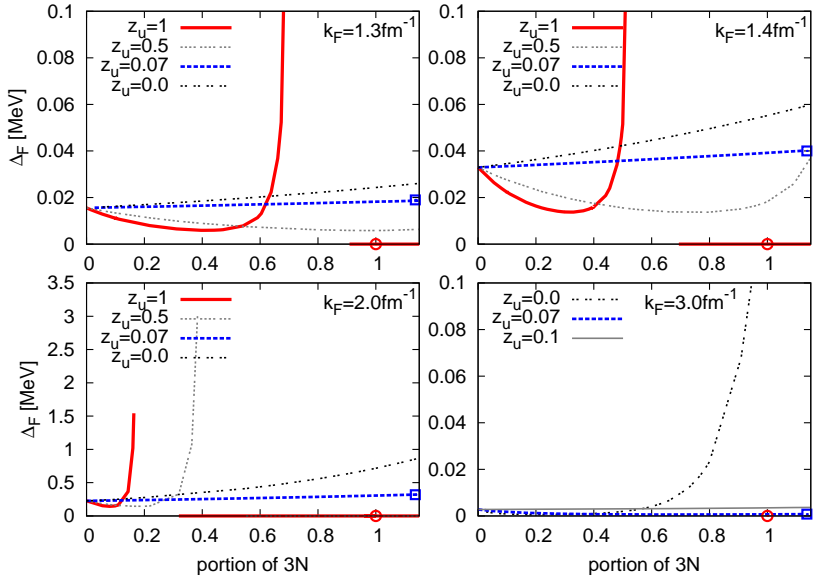}
\end{center}
\caption{Variation of the $^3P_2$ pairing gap with the strength of (i) 
the 3N Urbana-type force and of (ii) the relative strengths of its 
repulsive and attractive components, for the indicated values of $k_F$.  
The coupling constant $A_{2\pi}$ is scaled by the factor on the $x$-axis. 
The ratio $U_0/A_{2\pi}^{\mathrm{original}}$ is further scaled by the 
factor $z_u\leq 1$.  The original UIX potential corresponds to the portion 
of the 3N force equal to 1 and to $z_u=1$. The modified UIX potential 
corresponds to a portion of the 3N force equal to 1.1365 and $z_u=0.07$ 
(round points). 
(Color figure online.)
\label{Fig:DeltaVsNNN}
} \end{figure}

We do so first in Fig.~\ref{Fig:DeltaVsNNN} for four representative values 
of $k_F$, as indicated, and for different relative strengths 
$z_u=-U_0/A_{2\pi}^{\mathrm{original}}$.  As we gradually turn on the 
{\it original} 3N UIX interaction (cf.\ $z_u=1$ lines), the gap drops 
somewhat, and after a dramatic discontinuity it vanishes.  The results 
with the original UIX interaction fully turned on are indicated with 
circles.  For $k_F=3.0$~fm$^{-1}$, the gap vanishes already for a very 
weak 3N force.  If we now neglect the $U_0$ term (cf.\ $z_u=0$ lines), 
the gap in fact increases, as anticipated already from the comparison 
in Fig.~\ref{Fig:3N}.  For values of $U_0$ in between, the behavior 
again depends strongly on the details.  We attempt to elucidate the 
origin of the discontinuity in Sec.~\ref{Sec:NNNV}. 

Figure~\ref{Fig:DeltaVsNNN} includes results corresponding to the modified 
UIX parameters ($z_u=0.07$).  As we turn on the modified UIX force, the 
results trace the blue dashed lines.  The results with the modified UIX 
fully turned on are indicated with squares.  We observe that the modified 
UIX parameters are such that the inclusion of all, part, or none of 
it has a very weak influence on the gap in the uncoupled $^3P_2$ channel. 

\subsection{In-medium effects\label{Sec:NNNmed}} 

Above we have considered a Fermi-gas single-particle spectrum for 
the energy $\epsilon({\vec k'})$ entering the denominator of the gap 
equation (\ref{Eq:GapEq}), with no modification of the nucleon mass in 
the medium.  Let us now take into account a reduction of the nucleon 
effective mass, as per the results of Ref.~\cite{ZCL2008}, determined 
both with and without a 3N interaction, i.e., $m^{\ast}/m=0.82$ and $0.86$, 
respectively.  We now consider the solutions of the BCS equations upon 
replacing $m$ with $m^{\ast}$.  
%#
For this
%#
we focus on the near-peak solution, setting $k_F=2$~fm$^{-1}$. 

\begin{figure}
\begin{center}
\includegraphics[%
  width=0.55\linewidth,angle=-90,
  keepaspectratio]{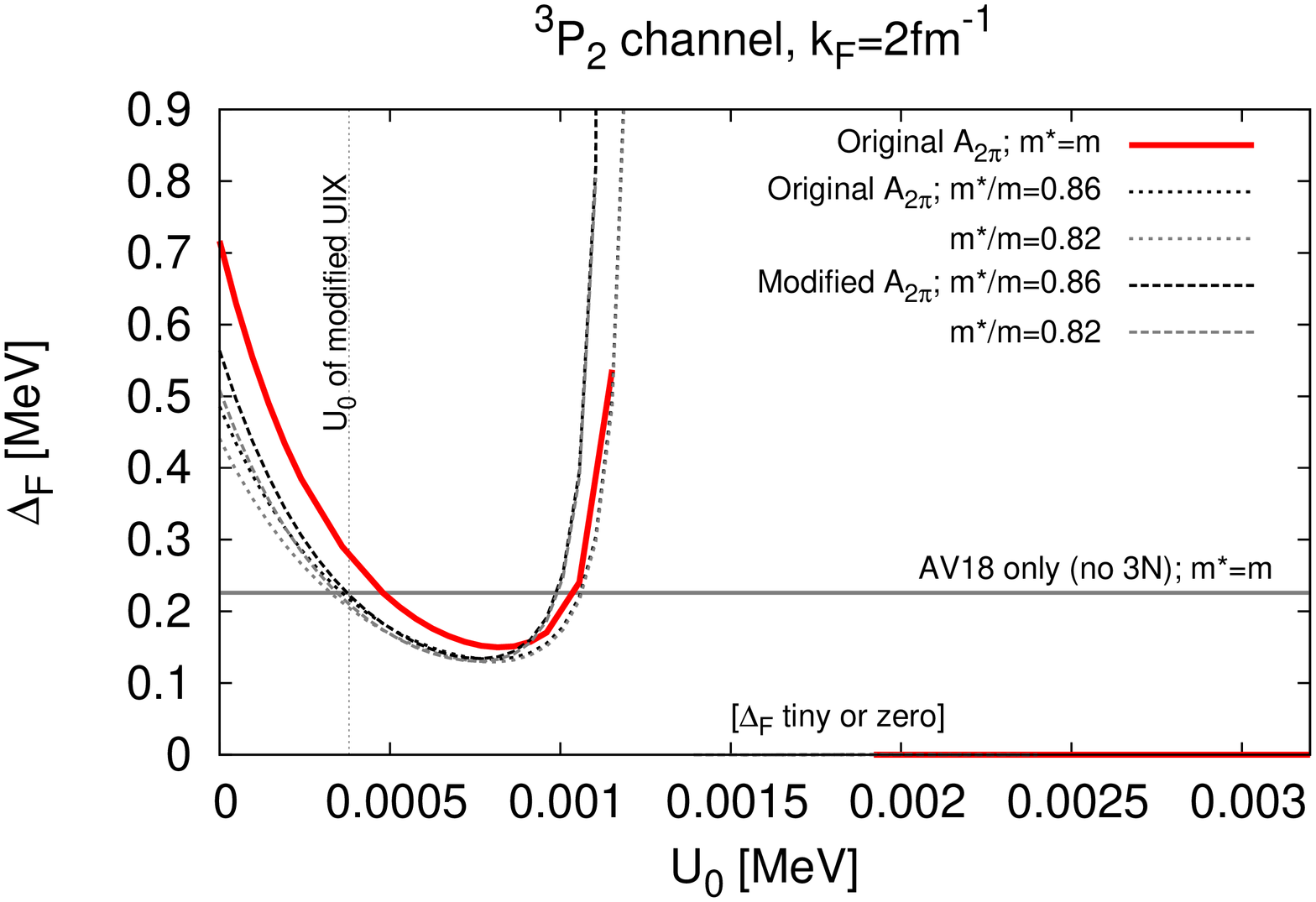}
\end{center}
\caption{Pairing gap in the $^3P_2$ channel in neutron matter for
$k_F=2$~fm$^{-1}$, calculated with the Argonne $v_{18}$ two-nucleon 
potential and an Urbana-type three-nucleon potential of varying 
strength, assuming different values for the nucleon effective mass. 
The original and modified UIX parameters are given in Table~\ref{table}. 
(Color figure online.)}
\label{Fig:EffMass}
\end{figure}

Keeping $A_{2\pi}$ fixed at its original or modified value, 
Fig.~\ref{Fig:EffMass} tracks the variation of the gap as the repulsive 
term $U_0$ is turned on for different $m^{\ast}/m$ values.  As in
the 2N-force results of Fig.~\ref{Fig:Coupled} for the uncoupled 
$^3P_2$ channel, it is found that the influence of the effective 
mass is quite weak.  Again we notice that the modified UIX potential 
has the curious effect of making the $^3P_2$ gap almost insensitive 
to the inclusion of the 3N force.  Whether this result originates 
in deeper consistency relations of the theories involved or is 
simply fortuitous is an open question.  In Sec.~{\ref{Sec:NNNV}} we 
will return to this point when dicussing the functional form of the 
total potential. 

The 3N force of the Urbana form can be strongly repulsive at short distances 
and hence may require careful treatment of short-range correlations.  To 
account for this effect, we simply introduce a defect function as explained 
in Sec.~\ref{Sec:Eqs}.  For the range $b$ we consider the values $0.6$ and 
$0.8$~fm. 

\begin{figure}
\begin{center}
\includegraphics[%
  width=0.55\linewidth,angle=-90,
  keepaspectratio]{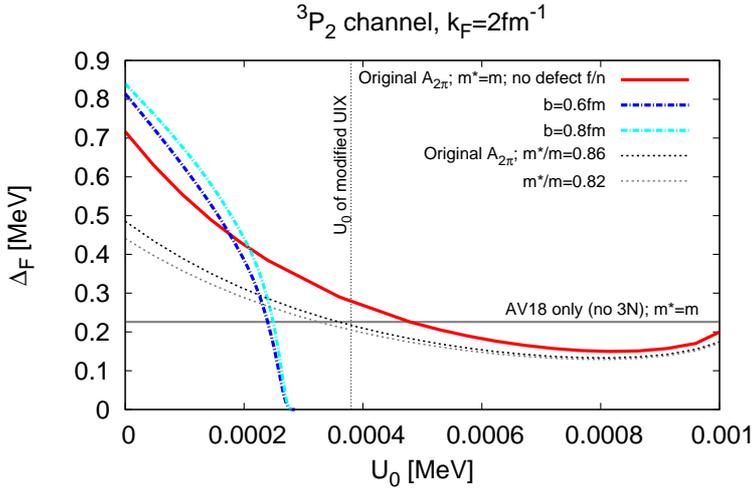}
\end{center}
%#
\caption{Pairing gap in the $^3P_2$ channel in neutron matter at 
$k_F=2$~fm$^{-1}$, calculated with the Argonne $v_{18}$ two-nucleon 
potential and a three-nucleon potential of Urbana IX type with the original attractive part and a 
varying repulsive strength (cf.\ Fig.~\ref{Fig:EffMass}), screened
by a defect function.  The gap is shown as a function of the coupling 
constant $U_0$ of the repulsive component. (Color figure online.)}
\label{Fig:DefectFunction}
\end{figure}

Results obtained with this simple procedure and for the original value of 
$A_{2\pi}$ are shown in Fig.~\ref{Fig:DefectFunction} as a function of $U_0$ 
(dash-dotted lines). For comparison and reference, the figure includes results 
from Fig.~\ref{Fig:EffMass}, and for clarity it zooms in on the domain of $U_0$ 
values for which the new results do not vanish. We observe that for small 
values of $U_0$ the effect of the correlations is rather weak.  Qualitatively, 
the main effect is the elimination of the discontinuity.  Already for small 
values of $U_0$ compared to its original value, the gap vanishes.

\subsection{The role of $V(k,k_F)$ \label{Sec:NNNV}}

\begin{figure}
\begin{center}
\includegraphics[%
  width=0.55\linewidth,angle=-90,
  keepaspectratio]{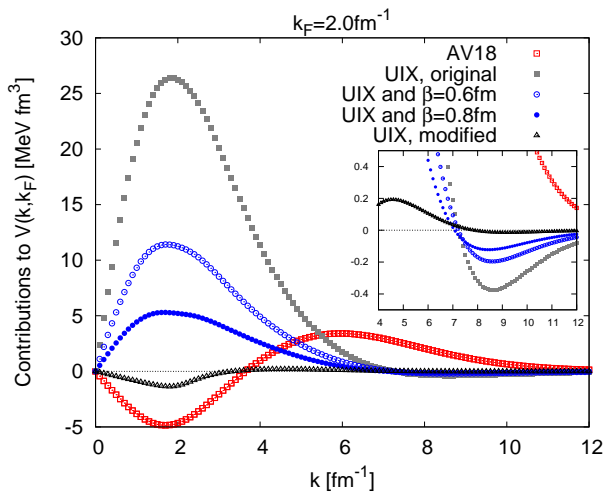}
\end{center}
\caption{Contributions to the total $V(k,k'=k_F)$ entering the gap equations in the $^3P_2$ channel, coming from the pure 2N AV18 potential and from variants of the 3N potential used in this work. 
 (Color figure online.)
\label{Fig:Pot2}} 
\end{figure}

To a large extent, the qualitative differences among the results obtained 
with the different implementations of 3N potentials can be traced back to the 
functional form of the potential itself.  To clarify this point, we plot in 
Fig.~\ref{Fig:Pot2} the 2N AV18 pairing potential $V (k,k_F)$ along with 
different possibilities we have considered for the 3N contribution 
$V^{(3)}(k,k_F)$ of Eq.~(\ref{Eq:V3k}), in the representative case 
$k_F=2$~fm$^{-1}$.  Specifically, for $V^{(3)}$ we have considered
(i) the original UIX potential, (ii) the 
%#
original UIX potential together with a defect function having range 
%#
$\beta=0.6$ or $0.8$~fm, and (iii) the modified UIX 
potential. (See Eqs.~(\ref{Eq:V3k}), (\ref{Eq:V3split}), (\ref{Eq:V3def}), 
and Table~\ref{table}.)  First, we note that the main contributions to 
the integrals in Eqs.~(\ref{Eq:Delta1})and (\ref{Eq:Delta2}) come from 
the neighborhood of $k'\approx k_F$.  It is then readily seen that 
Eq.~(\ref{Eq:Delta2}) generally has no solution if $V_F>0$.  One may 
also note that the modified UIX potential is negative around $k_F=2$~fm, 
thus enhancing the attraction contributed by the AV18 potential.  The 
original UIX potential, on the other hand, yields an overwhelmingly 
repulsive contribution, even when a defect function is introduced.  In 
order for the AV18 potential to compensate for the repulsion of the 
original UIX potential, the latter must be scaled by a factor 
$|V^{(AV18)}(k_F)|/V^{(3)}(k_F)\approx 0.2$. This is in line with 
what we observe in Fig.~\ref{Fig:DeltaVsNNN}: the gap at $k_F=2.0$~fm 
is finite as long as the 3N contribution is scaled by less than 0.2. 
As the scaling factor for UIX approaches that critical value from below, 
i.e., as $V_F$ approaches zero from below, only a very large value 
of $\Delta_F$ can ensure a solution of Eq.~(\ref{Eq:Delta2}) -- 
hence the steep rise in $\Delta_F$. Once $V_F$ turns positive, no 
solution is possible. 

In the case where a defect function is introduced, we infer from 
Fig.~\ref{Fig:Pot2} that, depending on $\beta$, a scaling factor between 
0.5 and 1.0 would ensure a negative $V_F$ for $k_F\approx 2$fm$^{-1}$.  
We may also infer from 
%#
the individual parameter values ($V^{(A_{2\pi})}_{F}=-2.7$, $-0.3$~MeV~fm$^3$) 
and ($V^{(U_0)}_{F}=13.9$, $7.5$~MeV~fm$^3$) for ($\beta=0.6$, $0.8$~fm), 
respectively, that a value of $U_0$ scaled by a factor $0.6$ to $0.8$ relative
to the original value could make the overall potential repulsive and kill 
the gap for $k_F\approx 2$fm$^{-1}$. 
As we observe in Fig.~\ref{Fig:DefectFunction}, the gap vanishes 
smoothly for a  
%#
lower scaling factor, i.e., before such a critical value of $V_F$ is reached; 
%#
hence the disappearence of the discontinuity. 

In closing, we may remark that the modified UIX potential seems to act as 
a small correction to the AV18 potential.  This is certainly not the case 
for the original UIX potential, even if we take into account certain 
in-medium effects. 

\section{Summary and perspectives \label{Sec:End}}

When exploring the nature of pairing phenomena in dense nuclear matter 
as it is likely to be present in the hadronic region of neutron star cores,
the irreducible three-nucleon interaction cannot be ignored.  Yet our 
knowledge of this interaction, as constrained by empirical data, is 
quite uncertain. The problem is aggravated by inconsistencies in 
competing microscopic many-body methods, which have never been tested
at the highest densities involved.  Consequently, the very existence of 
a triplet-$P$ pairing gap cannot be decided by theory at this point.
Hopefully, vital clues can be gleaned from the growing body of relevant
evidence from cooling curves of young neutron stars.

The situation described above emerged very clearly in the present study, 
wherein the gap equation for the triplet-$P$ channel in neutron matter 
was solved for the AV18 two-nucleon potential and two alternative UIX 
three-nucleon interactions.  We have shown that two parameterizations of 
the Urbana 3N force, both consistent with empirical saturation properties 
of nuclear matter within generally accepted many-body computational 
methods, 
%#
are likely to give
%#
conflicting answers to the binary question 
of whether there is a finite triplet$-P$ gap.  We have shown that a 
small fraction of the phenomenological repulsion of the 
%#
original
%#
UIX interaction suffices to kill the gap and that the introduction of 
short-range correlations does not guarantee reversal of this behavior.  
We cannot exclude the possibility that a special, even pathological, 
feature of the Urbana-type potential is responsible for our findings.  
Accordingly, we intend to investigate this issue of the existence 
of triplet-$P$ pairing for a broader selection of proposed 3N 
interactions.  It also remains to be seen whether the inclusion 
of coupling to the $F$ channel affects the conclusions drawn from 
the present study.  In particular, it is important to determine 
whether the sensitivity to the precise treatment of the 3N force 
persists for other candidate interactions.  At any rate, systematic 
and consistent studies are needed.  Their implications will lead 
to more meaningful comparisons with observational data and to a more 
profound understanding of nucleonic dynamics in the nuclear medium. 

\begin{acknowledgements}
The work of PP is supported by the Rare Isotope Science Project of the 
Institute for Basic Science, funded by the Ministry of Science, ICT and 
Future Planning and the National Research Foundation (NRF) of Korea 
(2013M7A1A1075764).  JWC is pleased to acknowledge research support 
from the McDonnell Center for the Space Sciences.  He is also grateful 
to the University of Madeira and its Center for Mathematical sciences 
for gracious hospitality during periods of extended residence.
\end{acknowledgements}

\pagebreak

%\bibliography{../../BIBS/NNN,../../BIBS/EFT}

\end{document}